\documentclass[usenatbib]{mn2e} \usepackage{psfig}\usepackage{graphicx}\usepackage{amsmath}\usepackage{amssymb}\usepackage{rotating}\usepackage{lscape} \usepackage{psfig}\usepackage{graphicx}\usepackage{amsmath}\usepackage{amssymb}\usepackage{rotating}\usepackage{lscape}

\title[XMM-Newton discovery of transient X-ray pulsar in NGC 1313]{XMM-Newton discovery of transient X-ray pulsar in NGC 1313}

\author[S.\,P.\,Trudolyubov] {S.\,P.\,Trudolyubov\thanks{E-mail: strudolyubov@gmail.com}\\ Institute of Geophysics and Planetary Physics, University of California, Riverside, CA 92521, USA\\}

\date{}

\pagerange{\pageref{firstpage}--\pageref{lastpage}} \pubyear{2008}

\begin{document}
\maketitle

\label{firstpage}

\begin{abstract}
We report on the discovery and analysis of the transient X-ray pulsar XMMU J031747.5-663010 detected in the 2004 
November 23 {\em XMM-Newton} observation of the spiral galaxy NGC 1313. The X-ray source exhibits pulsations with 
a period P$\sim$765.6 s and a nearly sinusoidal pulse shape and pulsed fraction $\sim$38\% in the 0.3-7 keV energy 
range. The X-ray spectrum of XMMU J031747.5-663010 is hard and is well fitted with an absorbed simple power law of 
photon index $\Gamma \sim 1.5$ in the 0.3-7 keV energy band. The X-ray properties of the source and the absence of 
an optical/UV counterpart brighter than 20 mag allow us to identify XMMU J031747.5-663010 as an accreting X-ray 
pulsar located in NGC 1313. The estimated absorbed 0.3-7 keV luminosity of the source L$_{\rm X}\sim 1.6\times 10^{39}$ 
ergs s$^{-1}$, makes it one of the brightest X-ray pulsars known. Based on the relatively long pulse period and 
transient behaviour of the source, we classify it as a Be binary X-ray pulsar candidate. XMMU J031747.5-663010 is 
the second X-ray pulsar detected outside the Local Group, after transient 18 s pulsating source CXOU J073709.1+653544 
discovered in the nearby spiral galaxy NGC 2403.   
\end{abstract}


\section{Introduction}
Since their discovery \citep{Giacconi71,Tananbaum72}, accreting X-ray pulsars have been major objects for both observational 
and theoretical study \citep{WSH83,Nagase89,Bildsten97}. The majority of known X-ray pulsars are high-mass binary systems 
with supergiant or Be donors, clearly associated with younger stellar populations and regions of recent star formation \citep{CC06}. 

Traditionally, the study of X-ray pulsars was limited to our Galaxy and neighboring Magellanic Clouds, because of the limited 
sensitivity and spatial resolution of previous X-ray missions. The advanced capabilities of a new generation of X-ray telescopes 
({\em Chandra} and {\em XMM-Newton}) have not only caused a rush of new X-ray pulsar discoveries in the Galaxy and Magellanic 
Clouds \citep{Lutovinov05,Chernyakova05,Karasev08,HP04,Edge04,McGowan07}, but also opened the possibility to extend X-ray pulsar 
search to more distant galaxies both inside \citep{O01,T05,TP08} and beyond the Local Group \citep{TPC07}. The nearby galaxies 
with recent star formation are especially suitable candidates for such studies, since they provide the environment in which X-ray 
pulsars are expected to be plentiful.
 
The nearby SB(s)d spiral galaxy NGC 1313 at 4.1 Mpc \citep{Mendez02}, provides an excellent opportunity to study X-ray 
source populations in a normal galaxy. The overall properties of NGC 1313 are similar to that of irregular Magellanic-type 
galaxies, and late-type spirals with vigorous recent and ongoing star formation. NGC 1313 was a target of X-ray observations 
with the {\em Einstein} (Fabbiano \& Trinchieri 1987), {\em ROSAT} (Colbert et al. 1995; Miller et al. 1998; Schlegel et al. 
2000), {\em ASCA} (Petre et al. 1994), {\em Chandra} (Schlegel et al. 2004; Zampieri et al. 2004), {\em XMM-Newton} (Miller et 
al. 2003; Schlegel et al. 2004; Smith et al. 2007) and {\em Suzaku} (Mizuno et al. 2007) observatories. These observations 
uncovered a substantial X-ray source population with three ultraluminous X-ray sources (two accreting binaries and the Type IIn 
supernova SN 1978K) among them.

In this paper, we report on the discovery of the coherent 765.6 s pulsations in the flux of transient X-ray source 
XMMU J031747.5-663010 in NGC 1313, using archival data of {\em XMM-Newton} observations. We study X-ray spectral properties of 
the source, search for its optical/UV counterparts and discuss its nature.

\section{Observations and data reduction} 
To study timing and spectral characteristics of XMMU J031747.5-663010, we used the data of 2004 November 23 {\em XMM-Newton} 
observation of NGC 1313 field with three European Photon Imaging Camera (EPIC) instruments (MOS1, MOS2 and pn) 
\citep{Turner01,Strueder01}, and the Optical Monitor (OM) telescope (Mason et al. 2001)(Table \ref{obslog}). We also used 
several 2000-2006 {\em XMM-Newton} \citep{Smith07} and 2002-2003 {\em Chandra} observations of the same field to obtain upper 
limits on the source flux when the source was not detected.

We reduced {\em XMM} data using {\em XMM-Newton} Science Analysis System (SAS v 7.0.0)\footnote{See http://xmm.vilspa.esa.es/user}. 
Before generating X-ray images, and source spectra and lightcurves, we performed standard screening of the original event 
files to exclude time intervals with high background levels, applying an upper count rate threshold of 20\% above average 
background level. The standard SAS tool {\em barycen} was used to perform barycentric correction on the original EPIC event 
files used for timing analysis. 

We generated EPIC-pn and MOS images of NGC 1313 field in the 0.3-7.0 keV energy band, and used the SAS standard maximum 
likelihood (ML) source detection script {\em edetect\_chain} to detect and localize point sources. We used bright X-ray 
sources with known counterparts from USNO-B catalog \citep{Monet03} and {\em Chandra} source lists to correct EPIC image 
astrometry. The astrometric correction was also applied to the OM images, using cross-correlation with USNO-B catalog. 
After correction, we estimate residual systematic error in the source positions to be of the order $\sim 1\arcsec$ for both 
EPIC and OM.

To extract EPIC-pn source lightcurves and spectra during the 2004 November 23 {\em XMM-Newton} observation, we used the 
elliptical region with semi-axes of 22$\arcsec$ and 18$\arcsec$ and position angle of 40$^{\circ}$. Due to the source 
proximity to the edge of EPIC-MOS CCD, the source counts were extracted from the elliptical region with semi-axes of 
20$\arcsec$ and 16$\arcsec$, including $\sim 70\%$ of the source energy flux. The adjacent source-free regions were used 
to extract background spectra and lightcurves. The source and background spectra were then renormalized by ratio of the 
detector areas. For spectral analysis, we used data in the $0.3 - 7$ keV energy band. In this analysis we use valid pn 
events with pattern 0-4 (single and double) and pattern 0-12 (single-quadruple) events for MOS cameras. To synchronize both 
source and background lightcurves from individual EPIC detectors, we used the identical time filtering criteria based on 
Mission Relative Time (MRT), following the procedure described in \cite{Robin_timing}. The background lightcurves were not 
subtracted from the source lightcurves, but were used later to estimate the background contribution in the calculation of 
the source pulsed fractions.

The EPIC spectra were grouped to contain a minimum of 20 counts per spectral bin in order to allow $\chi^{2}$ statistics, 
and fit to analytic models using the XSPEC v.12\footnote{http://heasarc.gsfc.nasa.gov/docs/xanadu/xspec/index.html} fitting 
package \citep{arnaud96}. EPIC-pn, MOS1 and MOS2 spectra were fitted simultaneously, but with normalizations varying 
independently. For timing analysis we used standard XANADU/XRONOS v.5\footnote{http://heasarc.gsfc.nasa.gov/docs/xanadu/xronos/xronos.html} 
tasks.

The data of {\em Chandra} observations was processed using the CIAO v3.4\footnote{http://asc.harvard.edu/ciao/} threads. We 
performed standard screening of the {\em Chandra} data to exclude time intervals with high background levels. For each 
observation, we generated X-ray images in the 0.3-7 keV energy band, and used CIAO wavelet detection routine {\em wavdetect} 
to detect point sources.

To estimate upper limits on the quiescent source luminosities, the {\em Chandra}/ACIS and {\em XMM}/EPIC count rates were 
converted into energy fluxes in the 0.3-7 keV energy range using Web PIMMS\footnote{http://heasarc.gsfc.nasa.gov/Tools/w3pimms.html}, 
assuming an absorbed power law spectral shape with photon index $\Gamma = 1.5$ and Galactic foreground absorbing column 
N$_{\rm H}$=$3.6\times10^{20}$ cm$^{-2}$.

In the following analysis we assume a distance of 4.1 Mpc for NGC 1313 \citep{Mendez02}. All parameter errors quoted 
are 68\% ($1\sigma$) confidence limits.

\section{Results}
\subsection{Source detection and optical counterparts}
\begin{figure}
\begin{tabular}{c}
\psfig{figure=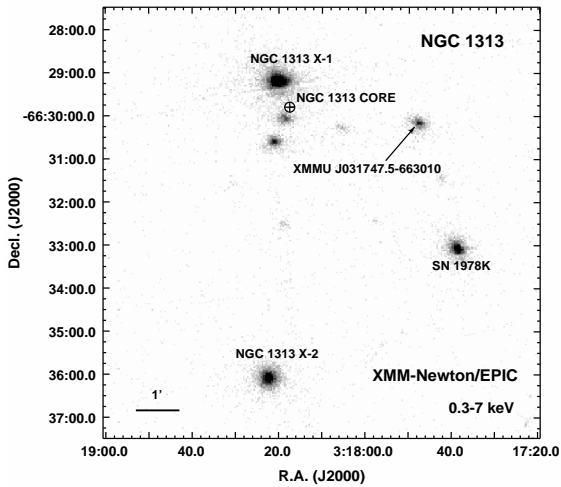,width=\columnwidth,angle=0.}\\
\\
\\
\psfig{figure=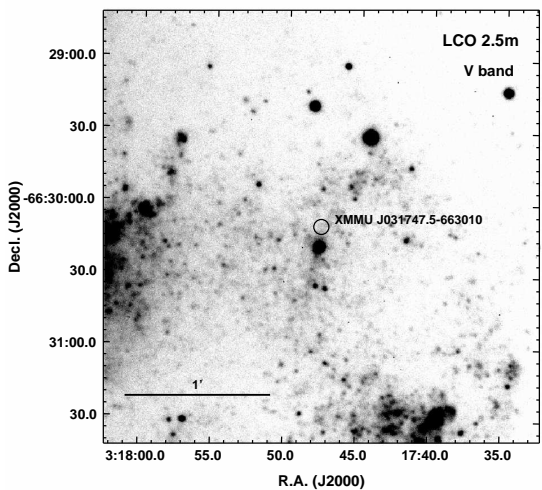,width=\columnwidth,angle=0.}
\end{tabular}
\caption{(Upper panel) Combined 0.3-7 keV {\em XMM}/EPIC image covering a 10$\arcmin\times10\arcmin$ region of NGC 1313, 
taken on Nov. 23, 2004. The position of new pulsar XMMU J031747.5-663010 is marked with an arrow, and the position of 
NGC 1313 nucleus \citep{Ryder95} is shown with a cross. Three other bright X-ray sources in the field are also marked for 
the reference \citep{Schlegel00}. (Lower panel) Optical V band image of NGC 1313 disk taken with Las Campanas Observatory 
2.5m du Pont telescope \citep{Kuchinski00}. The image is a 3$\arcmin\times3\arcmin$ square centered on the 
XMMU J031747.5-663010 position, shown with black circle of $3\arcsec$ radius ($3\sigma$).}
\label{image_general}
\end{figure}

A new X-ray source XMMU J031747.5-663010 has been discovered in the data of the 2004 November 23 {\em XMM-Newton} observation of 
the NGC 1313 field (Table \ref{obslog}). The estimated source luminosity was $\sim$1.6$\times 10^{39}$ ergs s$^{-1}$, assuming 
the distance of 4.1 Mpc. We measure the position of XMMU J031747.5-663010 to be $\alpha = 03^{h} 17^{m} 47.59^{s}, 
\delta = -66^{\circ} 30\arcmin 10.2\arcsec$ (J2000 equinox) with an uncertainty of $\sim 1.0\arcsec$ (Fig. \ref{image_general}). 
The projected galactocentric distance of XMMU J031747.5-663010 is $\sim 3\arcmin$ or $\sim 3.6$ kpc at 4.1 Mpc. The analysis of 
other archival observations of the same field with {\em XMM-Newton} and {\em Chandra} did not yield source detection with an 
upper limit (2$\sigma$) ranging from $\sim 2\times 10^{36}$ to $\sim 2\times 10^{37}$ ergs s$^{-1}$ (or $\sim$ 80-800 times 
lower than outburst luminosity), depending on the duration of the observation and instrument used (Table \ref{obslog}).  

The search for the optical counterparts using the deep images of NGC 1313 from Las Campanas Observatory 2.5m du Pont telescope 
\citep{Kuchinski00} did not yield stellar-like objects brighter than $\sim 21$ mag in V and $\sim 20$ mag in B band within 
the 3$\sigma$ error circle of XMMU J031747.5-663010. We also used the data of the 2004 November 23 {\em XMM-Newton}/OM observation 
to search for optical/UV counterparts to the source during its X-ray outburst (Fig. \ref{image_general}). We did not detect any 
stellar counterparts to XMMU J031747.5-663010 in the OM images down to the limit of $\sim 20$ mag in the V and U bands.

\subsection{X-ray pulsations}
\begin{figure}
\psfig{figure=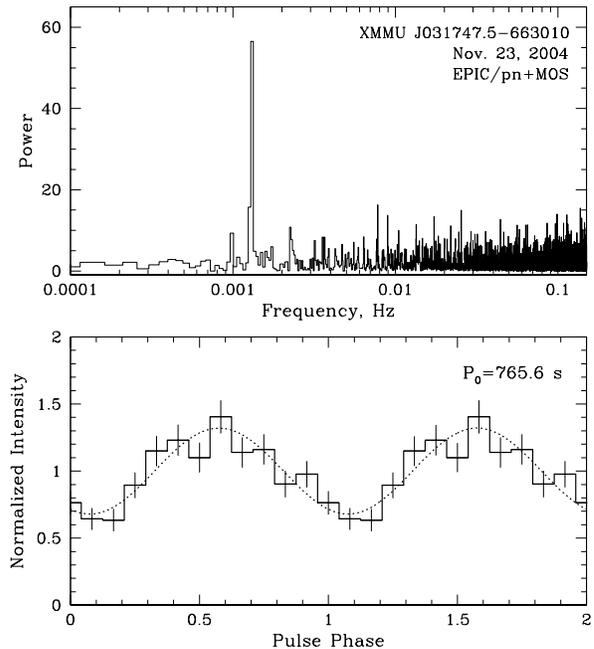,width=9.0cm,angle=0.}
\caption{(Upper panel) Power spectrum of XMMU J031747.5-663010 obtained using the data of 2004 Nov. 23 {\em XMM-Newton}/EPIC 
(EPIC-pn, MOS1 and MOS2 detectors combined) observation in the 0.3-7 keV energy band. (Lower panel) Corresponding background-corrected 
source X-ray pulse profile folded with most likely pulsation period (765.6 s). The sinusoidal fit to the pulse profile is shown for 
comparison with the dotted line.}
\label{pds_efold}
\end{figure}

We performed timing analysis of XMMU J031747.5-663010 using the 2004 November 23 data from all three {\em XMM-Newton}/EPIC 
detectors in the 0.3-7 keV energy band. After a barycentric correction of the photon arrival times in the original event lists, 
we performed a Fast Fourier Transform (FFT) analysis using standard XRONOS task {\em powspec}, in order to search for coherent 
periodicities. For the analysis of {\em XMM-Newton} data, we used combined synchronized EPIC-pn and MOS lightcurves with 2.6 s 
time bins to improve sensitivity. We found strong peak in the Fourier spectrum at the frequency of $\sim$1.3$\times10^{-3}$ Hz 
(Fig. \ref{pds_efold}, {\em upper panel}). The strength of the peak in the Fourier spectrum corresponds to the period detection 
confidence of $\sim 3\times10^{-9}$ \citep{Vaughan94}.

To estimate the pulsation period more precisely, we used an epoch folding technique, assuming no pulse period change during 
2004 November 23 observation. The most likely value of the pulsation period, 765.6 s (Table \ref{timing_spec_par}) was obtained 
fitting the peak in the $\chi^{2}$ versus trial period distribution with a Gaussian. The period errors in Table 
\ref{timing_spec_par} were computed following the procedure described in Leahy (1987). The source lightcurves were folded 
using the periods determined from epoch folding analysis. The resulting folded lightcurve in the 0.3-7 keV energy band 
during 2004 November 23 {\em XMM-Newton} observation is shown in Fig. \ref{pds_efold} ({\em lower panel}). The source 
has quasi-sinusoidal pulse profile in the 0.3-7 keV energy band with a pulsed fraction of $38\pm3$\%. The pulsed fraction 
was defined as (I$_{\rm max}$-I$_{\rm min}$)/(I$_{\rm max}$+I$_{\rm min}$), where I$_{\rm max}$ and I$_{\rm min}$ represent 
source intensities at the maximum and minimum of the pulse profile, excluding background photons.

\begin{figure}
\psfig{figure=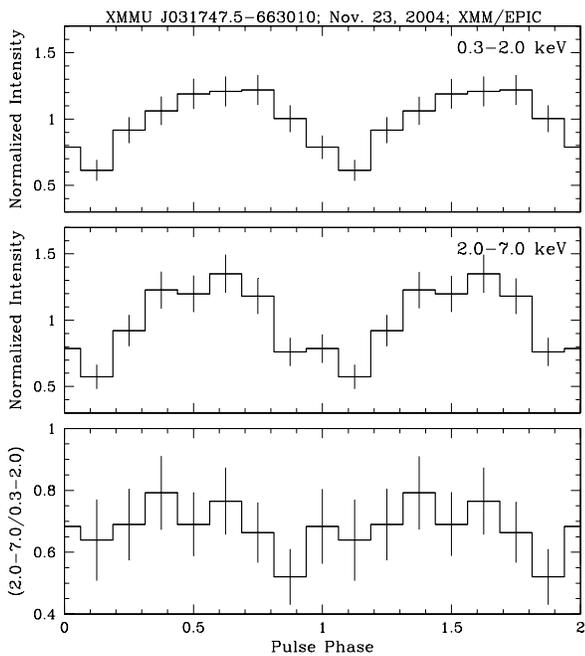,width=9.0cm,angle=0.}
\caption{Normalized X-ray lightcurves of XMMU J031747.5-663010 folded at the best pulsation period in the soft (0.3-2 keV, upper panel) 
and hard (2-7 keV, middle panel) energy bands along with hardness ratio (lower panel), computed taking background contribution into 
account.}
\label{mod_energy_depend}
\end{figure}

To investigate energy dependence of the source pulse profile, we extracted light curves in the soft (0.3-2 keV) and hard (2-7 keV) 
bands for 2004 Nov. 23 observation, and folded them at the corresponding best pulsation period (Fig. \ref{mod_energy_depend}). Both 
bands show quasi-sinusoidal pulse profiles. Because of the relatively poor statistics, we could detect only marginal difference 
between source pulse profiles at low and high energies, which have background-corrected pulsed fractions of 34$\pm$4\% and 40$\pm$4\%. 

\subsection{X-ray spectra}
The pulse phase averaged {\em XMM-Newton}/EPIC spectra of XMMU J031747.5-663010 can be adequately fit with the absorbed 
simple power law model with photon index, $\Gamma \sim 1.5$ and an equivalent hydrogen density N$_{\rm H}\sim2.3\times10^{21}$ 
cm$^{-2}$. The corresponding absorbed luminosity of the source in the 0.3-7 keV band is $\sim 1.6\times10^{39}$ ergs s$^{-1}$, 
assuming the distance of 4.1 Mpc. The best-fit spectral model parameters of the source are given in Table \ref{timing_spec_par}. 
The measured absorbing column $N_{\rm H}$ is $\sim$6 times higher than the Galactic hydrogen column in the direction of NGC 1313, 
3.6$\times10^{20}$ cm$^{-2}$ \citep{DL90}, consistent with an additional intrinsic absorption within the system and inside the 
disk of NGC 1313.

\begin{table*}
\caption{XMM-Newton and Chandra observations of NGC 1313 used in the analysis of XMMU J031747.5-663010. 
\label{obslog}}
\begin{tabular}{cccccccc}
\hline
Date, UT & Obs. ID  & Instrument & Mode/  & RA (J2000)$^{a}$ & Dec (J2000)$^{a}$ & Exp.$^{b}$ & Source luminosity$^{c}$\\
         &          &            & Filter & (h:m:s)          & (d:m:s)           & (ks)       & (ergs s$^{-1}$)\\
\hline
2000 Oct. 17 & 0106860101 & EPIC  & Full/Medium& 03:18:22.61 & -66:30:36.4 & 21.6(MOS)/21.6(pn) & $<4.1\times10^{36}$\\
2002 Oct. 13 & 2950       & ACIS-S& Very Faint & 03:18:32.00 & -66:31:10.0 & 19.9               & $<2.9\times10^{36}$\\
2002 Nov. 09 & 3550       & ACIS-I& Very Faint & 03:17:55.00 & -66:34:40.0 & 14.5               & $<4.0\times10^{36}$\\
2003 Oct. 02 & 3551       & ACIS-I& Very Faint & 03:17:55.00 & -66:34:40.0 & 14.8               & $<3.2\times10^{36}$\\\
2003 Dec. 21 & 0150280301 & EPIC  & Full/Thin  & 03:18:20.30 & -66:37:03.2 & 10.0(MOS)/8.4(pn)  & $<1.4\times10^{37}$\\
2003 Dec. 23 & 0150280401 & EPIC  & Full/Thin  & 03:18:19.69 & -66:37:02.0 & 7.5(MOS)/6.2(pn)   & $<1.8\times10^{37}$\\
2003 Dec. 25 & 0150280501 & EPIC  & Full/Thin  & 03:18:19.26 & -66:37:01.3 & 8.6(MOS)/7.0(pn)   & $<2.0\times10^{37}$\\
2004 Jan. 08 & 0150280601 & EPIC  & Full/Thin  & 03 18 16.60 & -66 36 56.1 & 10.9(MOS)/9.5(pn)  & $<1.4\times10^{37}$\\
2004 Jan. 16 & 0150281101 & EPIC  & Full/Thin  & 03:18:14.90 & -66:36:51.6 & 8.6(MOS)/7.0(pn)   & $<1.7\times10^{37}$\\
2004 Aug. 23 & 0205230401 & EPIC  & Full/Thin  & 03:18:31.90 & -66:35:33.1 & 12.1(MOS)/11.2(pn) & $<1.7\times10^{37}$\\
2004 Nov. 23 & 0205230501 & EPIC  & Full/Thin  & 03:18:25.09 & -66:36:59.5 & 15.6(MOS)/14.0(pn) & 1.6$\times10^{39}$ \\
2004 Jun. 05 & 0205230301 & EPIC  & Full/Thin  & 03:18:19.98 & -66:34:48.8 & 11.6(MOS)/10.0(pn) & $<1.1\times10^{37}$\\
2005 Feb. 07 & 0205230601 & EPIC  & Full/Thin  & 03:18:12.06 & -66:36:31.1 & 12.4(MOS)/10.9(pn) & $<1.4\times10^{37}$\\
2006 Mar. 06 & 0301860101 & EPIC  & Full/Medium& 03:17:27.38 & -66:33:08.0 & 21.5(MOS)/19.9(pn) & $<7.1\times10^{36}$\\
2006 Oct. 15 & 0405090101 & EPIC  & Full/Thin  & 03:18:23.51 & -66:30:38.9 & 100.0(MOS)/98.0(pn)& $<2.3\times10^{36}$\\
\hline  
\end{tabular}

$^{a}$ -- pointing coordinates\\
$^{b}$ -- instrument exposure used in the analysis\\
$^{c}$ -- estimated luminosity of XMMU J031747.5-663010 in the 0.3-7 keV band\\ 

\end{table*}

\begin{table*}
\caption{X-ray pulsation parameters and spectral fit information for XMMU J031747.5-663010. 
\label{timing_spec_par}}
\begin{tabular}{cccccccccl}
\hline
\multicolumn{2}{c}{Timing Parameters}&\multicolumn{8}{c}{POWERLAW*WABS Spectral Model Parameters}\\
\hline
 Period & PF$_{0.3-7 keV}$ & N$_{\rm H}$                & Photon & Flux$_{\rm abs}^{b}$ & Flux$_{\rm abs.corr}^{c}$ & $L_{\rm abs}$$^{d}$& $L_{\rm abs.corr}$$^{e}$& $\chi^{2}$ & Instrument\\
  (s)   &       (\%)$^{a}$ &($\times 10^{20}$ cm$^{-2}$)& Index  &                     &                            &                    &                         & (d.o.f.)   & \\       
\hline
$765.6\pm4.0$ & $38\pm3$ & $23\pm4$ & $1.48^{+0.10}_{-0.09}$ & $7.84\pm0.35$ & $9.98^{+0.55}_{-0.52}$ & 1.58 & 2.01 & 64.9(52) & pn+M1+M2\\
\hline
\end{tabular}

$^{a}$ -- pulsed fraction in the $0.3-7$ keV energy band\\
$^{b}$ -- absorbed model flux in the $0.3 - 7$ keV energy range in units of $10^{-13}$ erg s$^{-1}$ cm$^{-2}$\\
$^{c}$ -- unabsorbed model flux in the $0.3 - 7$ keV energy range in units of $10^{-13}$ erg s$^{-1}$ cm$^{-2}$\\
$^{d}$ -- absorbed luminosity in the $0.3 - 7$ keV energy range in units of $10^{39}$ erg s$^{-1}$, assuming the distance of 4.1 Mpc\\
$^{e}$ -- unabsorbed luminosity in the $0.3 - 7$ keV energy range in units of $10^{39}$ erg s$^{-1}$, assuming the distance of 4.1 Mpc\\

\end{table*}

\begin{figure}
\psfig{figure=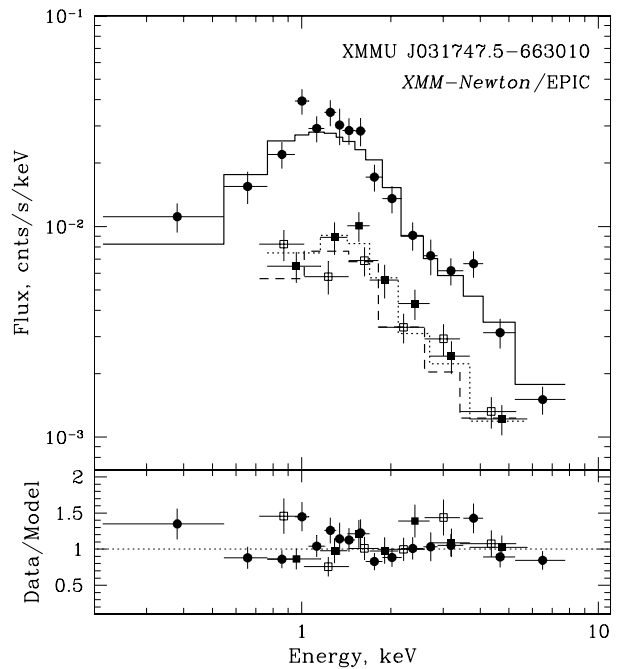,width=9.0cm,angle=0.}
\caption{EPIC count spectra and model ratios of XMMU J031747.5-663010 during the 2004 Nov. 23 observation. The EPIC-pn data is plotted 
with filled circles, while EPIC-MOS1 and MOS2 data is shown with filled and open squares respectively. The best-fit absorbed power law 
model approximation of EPIC-pn, MOS1 and MOS2 data is shown with solid, dotted and dashed histograms.}
\label{spec_fig}
\end{figure}

\section{Discussion}
The absence of the bright optical counterpart to XMMU J031747.5-663010, its overall X-ray properties (spectrum, pulsations, 
transient behaviour), and positional coincidence with NGC 1313 disk, allow us to conclude that it should be located outside 
our Galaxy and probably belongs to NGC 1313. The X-ray pulsations and energy spectrum of XMMU J031747.5-663010 imply that 
it is almost certainly an accreting highly-magnetized neutron star in a high-mass binary system (White, Swank \& Holt 1983; 
Nagase 1989). The association with NGC 1313 makes this source an extremely bright object with luminosity 
L$_{\rm X}\sim 1.6\times 10^{39}$ ergs s$^{-1}$, greatly exceeding the isotropic Eddington luminosity limit for a 
$1.4 M_{\odot}$ neutron star accreting hydrogen-rich material. 

The relatively long pulse period of XMMU J031747.5-663010 (765.6 s) places it among the systems with a companion that is 
either a supergiant or a Be star on a Corbet diagram \citep{Corbet86}. The transient behavior of the source lends support 
to the interpretation of this source as yet another Be binary, since the majority of Be systems display recurrent/transient 
outbursts. An extremely high luminosity of the source still falls into the luminosity range observed in the Be X-ray pulsars, 
with one system, A0538-66 known to reach similar luminosity during its giant (Type II) outburst \citep{WC78,Skinner82}. The 
high luminosity of XMMU J031747.5-663010 is also consistent with theoretical predictions for super-Eddington accretion onto 
highly magnetized ($B \gtrsim 10^{12}$ G) neutron star \citep{BS76}. 

The high-mass nature of the system implies its relatively young age, consistent with its location within one of the spiral 
arms of NGC 1313 (Fig. \ref{image_general}). Since the Be interpretation still remains preliminary, optical identification is 
essential to determine the nature of the system. For optical identification, deeper optical observations are needed. The 
follow-up monitoring observations with {\em Chandra} and {\em XMM-Newton} are needed to test if it shows recurrent outbursts. 
Future X-ray observations of XMMU J031747.5-663010, if it reappears, could improve source localization and study long-term 
evolution of its X-ray properties and X-ray pulsation.   

XMMU J031747.5-663010 is the second X-ray pulsar detected outside the Local Group, after 18 s pulsating X-ray source CXOU 
J073709.1+653544 in the nearby spiral galaxy NGC 2403 \citep{TPC07}. Both sources are extremely bright transient systems with 
total luminosities $\gtrsim 10^{39}$ ergs s$^{-1}$, and probably belong to a rare class of most luminous Be binary X-ray pulsars. 
Similar systems can be detected and studied effectively with a series of moderately deep (10-50 ks) monitoring {\em XMM-Newton} 
and {\em Chandra} observations up to the distances of several Mpc. Therefore, the detailed analysis of the existing archival and 
new observations of nearby spiral and irregular galaxies have a potential to significantly increase statistics of these systems 
and provide us with better understanding of their nature and connection to the underlying stellar population.  

\section*{Acknowledgments}
The author would like to thank the referee for comments and suggestions that improved the paper. This research has made use of 
data obtained through the High Energy Astrophysics Science Archive Research Center Online Service, provided by the NASA/Goddard 
Space Flight Center. {\em XMM-Newton} is an ESA Science Mission with instruments and contributions directly funded by ESA Member 
states and the USA (NASA). {\em Chandra} X-ray Observatory is operated by the Smithsonian Astrophysical Observatory on behalf of 
NASA. This research also made use of NASA/IPAC Extragalactic Database (NED), which is operated by the Jet Propulsion Laboratory, 
California Institute of Technology, under contract with NASA.

\end{document}